\documentclass[fleqn,twoside]{article}
\usepackage{espcrc2}
\usepackage{epsfig}
\usepackage{graphicx}
\usepackage[figuresright]{rotating}

\newcommand{\AmS}{{\protect\the\textfont2
  A\kern-.1667em\lower.5ex\hbox{M}\kern-.125emS}}

\title{The liquid Argon TPC: a powerful detector for future neutrino experiments and proton decay searches}

\author{A. Ereditato\address[INFN]{Istituto Nazionale di Fisica Nucleare, INFN Sezione di Napoli, Napoli, Italy}
                and
        A. Rubbia\address[ETHZ]{Institut f\"{u}r Teilchenphysik, ETHZ, CH-8093 Z\"{u}rich,
Switzerland}}
       
\begin{document}

\begin{abstract}
We discuss the possibility of new generation neutrino and astroparticle physics experiments 
exploiting the liquid Argon Time Projection Chamber (LAr TPC) 
technique, following a graded strategy that envisions applications with increasing detector masses
(from 100 ton to 100 kton). 
The ICARUS R\&D program has already demonstrated that the
technology is mature with the test of the T600 detector at surface.
Since 2003 we have been working with the conceptual design of a very large LAr TPC
with a mass of 50-100 kton to be built by employing a monolithic technology based on the use of industrial,
large volume, cryogenic tankers developed by the petro-chemical industry.
Such a detector, if realized, would be an ideal match for a 
Super Beam, Beta Beam or Neutrino Factory, 
covering a broad physics program that includes the detection of atmospheric, 
solar and supernova neutrinos, and searches for proton decay, in addition to the rich 
accelerator neutrino physics program.
A "test module" with a mass of the order of 10 kton operated underground 
or at shallow depth would represent a necessary milestone towards
the realization of the 100 kton detector, with an interesting physics program on its own.
In parallel, physics is calling for a shorter scale application of the LAr TPC technique
at the level of 100 ton mass, for low energy neutrino physics and for use as a near station
setup in future long baseline neutrino facilities.
We outline here the main physics objectives and the design of such a detector for operation 
in the upcoming T2K neutrino beam.
We finally present the result of a series of R\&D studies conducted with the aim of validating the design of the 
proposed detectors.

\vspace{1pc}
\end{abstract}

\maketitle

\section{Introduction}

The present generation of neutrino experiments will further clarify the scenario of neutrino mixing
by reducing the errors on the oscillation parameters, by confirming the $\nu_\mu\rightarrow \nu_\tau$ oscillation
channel and by proving or disproving the existence of a fourth and sterile neutrino.
This generation of experiments can be assumed to provide solid and convincing results by $\sim$ 2010. By then, in order to 
proceed along this fascinating line of research a new technological step will be required, both concerning the
neutrino beam facilities (intensity and reliability) and the detectors (performance and mass).
By that time the neutrino community will be confronted with the measurement of the (so far) unknown 
$\theta_{13}$ mixing angle. The relevance of this physics result is twofold: on the one hand, 
if this angle turns out to be non vanishing, the actual behavior of the three neutrino mixing scheme will be proved;
on the other hand, the measurement of $\theta_{13}$ 
will open the way to the future searches for a possible CP violation in the leptonic sector.
After one more decade one could assume that the latter research subject could be attacked by a further 
beam facility and detector generation, whose features can be hardly predicted today, apart from a generic request of 
unprecedented high beam intensity and huge detector mass.

Following the above (oversimplified) scenario, as far as detectors are concerned, it is clear that
neutrino physics is now calling for new devices able to meet the challenge of the high intensity, and also 
capable of improved particle identification and background rejection performance.
The large mass of these detectors and their complexity will obviously imply important financial investments.
This will make mandatory an appropriate and economical use of these facilities, by extending their application
to other fields, such as astroparticle physics and matter unstability searches.
The realization of these detectors will also imply (actually, already by now) a vigorous and coordinated effort of the community
on the required R\&D activities, as well as a wise exploitation of existing resources and infrastructure, seeking for 
complementarity of the different worldwide projects.

We believe that the detection technique of the liquid Argon Time Projection Chambers is well suited to answer
the demanding experimental requirements originating from the above considerations.
In the last few years, in particular, we have been proposing a global research strategy centered on
both the design and proposal of experiments exploiting this technique and on the related R\&D strategy~\cite{nuint04}. 
The main milestones of our line of thought can be summarized as follows:

\begin{itemize}

\item after more than 20 years of R\&D work from the ICARUS Collaboration all the basic elements of the LAr TPC technique
have been successfully developed;

\item no "black magic" is required to exploit the technique and realize detectors of increasing mass although
specific research work might be needed for advanced applications;

\item we stress the importance of small laboratory prototype detectors to get acquainted with the technological
issues and to study specific R\&D subjects;

\item the next-to-come physics application might likely be the realization of a $\sim$ 100 ton detector for a 
LBL neutrino facility, able to study low energy neutrino interactions, to ascertain the capability of the
technique in performing high precision neutrino physics experiments;

\item we have performed the conceptual design and preliminary engineering studies of a very large mass (50-100 kton)
LAr TPC facility for ultimate proton decay searches and precision studies of the neutrino mixing matrix;

\item the above large mass facility, representing more than a factor 100 mass increase with respect to
the ICARUS modular design approach, will require a two step strategy, possibly envisioning the realization of an "engineering module" with a mass
of the order of 1 kton (able to prove the scalability of the detector) and of a $\sim$ 10 kton detector able to provide
a rich physics program and to justify the further factor of 10 mass increase for the very large facility;

\item a complete R\&D program has been identified for several key issues. We already obtained interesting results and
more projects are underway or planned for the next future.

\end{itemize}

After a brief introduction on the LAr detection technique,
the above milestone are described in more detail in the next Sections together with the detector design features and 
experimental results from R\&D studies.

\section{The liquid Argon TPC technique}

The Liquid Argon Time Projection Chamber was conceived and
proposed by C.~Rubbia in 1977~\cite{intro1} as a tool for uniform and high accuracy imaging of massive detector volumes. 
The operating principle of the LAr TPC is based on
the fact that in highly purified LAr ionization tracks could be transported
undistorted by a uniform electric field over distances of the order of meters. 
Imaging is provided by wire planes  placed at the end of the drift path, continuously sensing and recording the 
signals induced by the drifting electrons. Liquid Argon is an ideal medium since it provides
high density, high ionization and scintillation yields, and is intrinsically safe and cheap. 

Non-destructive readout of ionization electrons by charge induction allows to detect the signal of electrons crossing
subsequent wire planes with different orientation. This provides several
projective views of the same event, hence allowing for space point reconstruction and precise calorimetric measurement.
The particle momentum can be inferred via a multiple scattering measurement, while
the detection of the local energy deposition ($dE/dx$) can provide
$e/\pi^0$ separation and particle identification through a range versus $dE/dx$ measurement.
The total energy reconstruction of the event is performed by charge integration within
the detector volume, being the detector a full-sampling, homogenous calorimeter.

The main technological challenges of this technology are summarized
elsewhere~\cite{t600paper} and included techniques of Argon purification,  operation of wire 
chambers in cryogenic liquid and without charge amplification, 
low-noise analog electronics, continuous wave-form recording and digital signal processing.
The extensive ICARUS R\&D program dealt with
studies on small LAr volumes, LAr purification methods, readout schemes and electronics, 
as well as studies with several prototypes of increasing mass on purification technology,
collection and analysis of physics events, long duration tests and
readout~\cite{3tons,Cennini:ha,50lt}. 

The realization of the 600 ton ICARUS detector (T600) culminated with the full test of one of the two 300 ton modules 
carried out at surface~\cite{t600paper}. This test demonstrated that
the LAr TPC technique can be operated at large mass scale with a drift length of 1.5~m.
Data taking with cosmic-ray events allowed to assess
the detector performance in a quantitative way~\cite{Amoruso:2003sw,Amoruso:2004dy,gg2,gg3,gg1}. 
Installation of the T600 module at the Gran Sasso Underground Laboratory is currently
on-going.

Nowadays, LAr TPC detectors can be readily used in a broad energy range, from MeV up to multi-GeV with
high event reconstruction efficiency. At the same time, low thresholds for particle identification
are possible thanks to the high granularity.  In addition to the natural use as underground facilities,
one can also operate the detectors at shallow depth owing, once more, to the high
granularity which permits the separation of signal from background. Finally, one can realistically think to embed a 
LAr TPC in a magnetic field for charge discrimination~\cite{Rubbia:2001pk}.
Implementations at different mass scales ($e.g.$ from 100~tons to 100~ktons) are therefore conceivable as well as
technically and economically sound.
In the following, we give some examples according to the global strategy outlined in the previous Section.

\begin{figure}[htb]
\centering
\epsfig{file=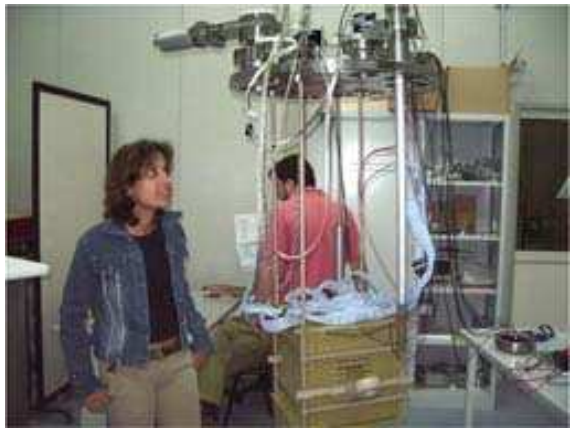,width=7.5cm}
\vspace{-1cm}
\caption{\small LAr TPC prototype operating at INFN Napoli.}
\label{fig:napolilab}
\end{figure}

\begin{figure}[htb]
\centering
\epsfig{file=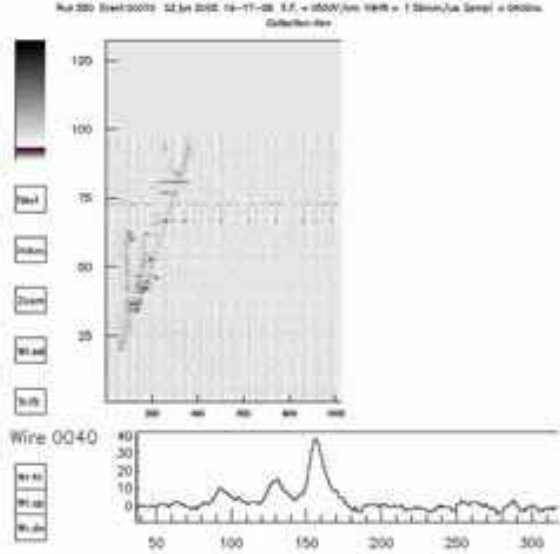,width=8.0cm}
\vspace{-1cm}
\caption{\small Display of a cosmic-ray event taken with the TPC prototype at INFN Napoli.}
\label{fig:eventnap}
\end{figure}

In parallel to the design of new experiments employing LAr TPC detectors,
one could readily profit of the know-how acquired on the technique to setup small laboratory prototype
detectors, able to permit the study of specific technological subjects and to gather the necessary expertise with all the 
related experimental issue.
As an example, Fig.~\ref{fig:napolilab} shows a test chamber we have 
been operating at INFN Napoli, in particular for the study 
of UV laser-liquid Argon calibration~\cite{napolipaper}.  
The display of a cosmic-ray event taken with this detector is shown in Fig.~\ref{fig:eventnap}.

\section {A large mass liquid Argon TPC detector with charge imaging and light readout}

A very large LAr TPC with a mass ranging from 50 to 100~kton would deliver extraordinary physics output, 
sometimes called ``megaton physics", owing to the excellent event reconstruction
capabilities provided by the technique.
Coupled to future Super Beams~\cite{Ferrari:2002yj}, Beta Beams or Neutrino Factories~\cite{Rubbia:2001pk,Bueno:2001jd}
it could greatly improve our understanding of the mixing matrix in the lepton sector with
the goal of measuring the CP-phase. At the same time, it would
allow to conduct astroparticle experiments of unprecedented sensitivity.

\begin{figure}[htb]
\centering
\epsfig{file=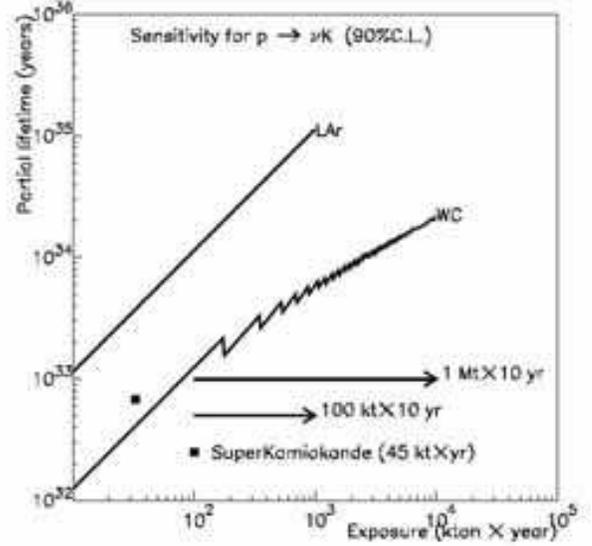,width=8cm}
\vspace{-1cm}
\caption{\small Comparison between the sensitivities of a 100 kton LAr TPC and a 650 kton water Cerenkov
in the search for proton decay:  p $\rightarrow$ $K$ + $\nu$.}
\label{fig:protondecay}
\end{figure}

Let us mention, as an example, the physics reach of such a detector for nucleon decay searches.
For the two prong decay mode: p $\rightarrow$ $K$ + $\nu$, a 100 kton LAr detector, thanks to its 
high detection efficiency ($>$ 95\%) and to the high background rejection power could effectively compete 
with a much larger mass (650 kton) water Cerenkov detector,
reaching a sensitivity limit of $\tau \sim 10^{35}$ years for a ten years run (Fig.\ref{fig:protondecay}).
This features strongly supports the concept of complementarity between the two detection methods for proton decay searches.

The basic features of the proposed detector~\cite{Rubbia:2004tz,nuint04,multimw,Rubbia:2004yq,nufact04}
can be summarized as follows.
The baseline design envisions a single 100 kton ``boiling'' cryogenic tanker at atmospheric
pressure for a stable and safe equilibrium condition, since temperature is constant while Argon is boiling.
The evaporation rate is small, less than $10^{-3}$ of the total volume per day, and is compensated
by refilling of the evaporated Argon volume. 
The detector signal is provided by charge imaging, scintillation and Cerenkov light readout,
for a complete and redundant event reconstruction. 
A peculiar feature is represented by the fact that
the detector is running in double-phase mode. In order to allow for drift lengths as long as $\sim$ 20 m,
which provides an economical way to increase the volume of the detector with a constant number
of channels, charge attenuation will occur along the drift due to attachment to the remnant impurities present
in the LAr. This effect can be compensated with charge amplification near the anodes located in the gas phase.

The cryogenics design of the proposed detector relies on the industrial
know-how in the storage of liquefied natural gases (LNG),
which developed in the last decades, driven by the petrochemical industry. 
The technical challenges associated to the design, construction and safe operation 
of large cryogenic tankers have already been addressed and engineering problems
have been solved. 
The current state-of-the-art contemplates cryogenic tankers of
200000~m$^3$. In the world presently exist $\sim$~2000 tankers
with volumes larger than 30000~m$^3$, with the majority built during the last 40 years. 
Technodyne International Limited, UK~\cite{Technodyne}, which has expertise in the design of LNG tankers, has produced
for us a feasibility study in order to understand and clarify all the issues related to the operation of a large
underground LAr detector. 

\begin{figure}[htb]
\centering
\epsfig{file=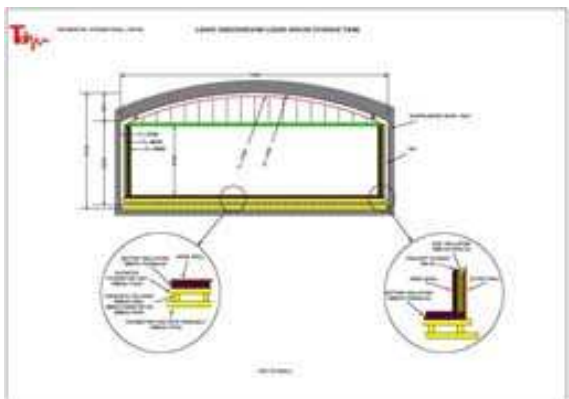,width=7.5cm}
\vspace{-1cm}
\caption{\small Conceptual design of a 100~kton liquid Argon cryogenic tanker developed
in collaboration with Technodyne International Limited.}
\label{fig:glacier}
\end{figure}

The outcome of the collaboration with Technodyne is shown in Fig.~\ref{fig:glacier} where the engineering design
of the large detector is presented.
The detector is characterized by the large fiducial volume of LAr included in a tanker with external dimensions
of approximately 40 m in height and 70 m in diameter. A cathode located at the bottom of the 
inner tanker volume creates a drift electric field of the order of 1~kV/cm over a distance of about 20~m. 
In this field configuration ionization electrons
are moving upwards while positive ions are going downward. The electric field is delimited on the sides of the tanker
by a series of ring electrodes (race-tracks) placed at the appropriate potential by a voltage divider.

The tanker contains both liquid and gas Argon phases at equilibrium. Since purity is a concern for very long
drifts of 20 m, we already mentioned that the inner detector could be operated in double-phase mode.
In order to amplify the extracted charge one can consider various options: proportional amplification
with thin readout wires, GEMs or LEMs (see Section~5). 
After a drift of 20 m at 1 kV/cm, the electron cloud diffusion reaches approximately a size of 3 mm, which corresponds to the
envisaged readout pitch. If one assumes that the electron lifetime is at least 2~ms~\cite{gg3},
one then expects an attenuation of a factor $\sim$ 150 over the distance of 20~m,
compensated by the proportional gain at the anodes.
We remind that the expected attenuation factor will not introduce
any detection inefficiency, given the value of $\sim$ 6000 ionization electrons per millimeter produced 
along a minimum ionizing track in LAr.
The number of readout channels is of the order of 100000 served by about 100 electronic crates placed on top of the tanker.

In addition to charge readout, one can envision to locate PMTs around the inner surface of the tanker. 
Scintillation and Cerenkov light can be readout essentially independently. 
LAr is a very good scintillator with a yield of about 50000 $\gamma$/MeV (at zero electric field). 
However, this light is distributed around a line at 128~nm and, therefore, 
a PMT wavelength shifter (WLS) coating is required. 
Cerenkov light from penetrating muon tracks has been successfully detected in a LAr TPC~\cite{gg2};
this much weaker radiation with about $700\ \gamma/$MeV produced between 
160~nm and 600~nm for an ultra-relativistic muon can be separately identified with
PMTs without WLS coating, since their efficiency for the DUV light will be very small. 
A total of 1000 8" PMTs would be required to readout the scintillation signal and about 27000 PMTs to 
cover 20\% of the inner tanker surface to possibly detect the Cerenkov signal.

The operation of such a large facility is certainly challenging, although solutions exist for the main issues, such as cryogenics,
provision of LAr, filling-up, purification, etc.~\cite{Rubbia:2004tz,nuint04,multimw,Rubbia:2004yq,nufact04}.
Here we only stress that an "in situ" plant producing liquid Argon might be an
economical solution, also considering the cost of 
transporting the liquid from a distant production site. Assuming a filling speed of 150 ton/day (compatible with 
the daily production rate of such a plant) one would need about two years to fill the detector.
Since the boiling-off of the tanker corresponds to 45 ton/day (for a 5 W/m$^2$ heat input) it would take nearly 10 years
to completely evaporate the whole detector volume.

\begin{figure}[htb]
\centering
\epsfig{file=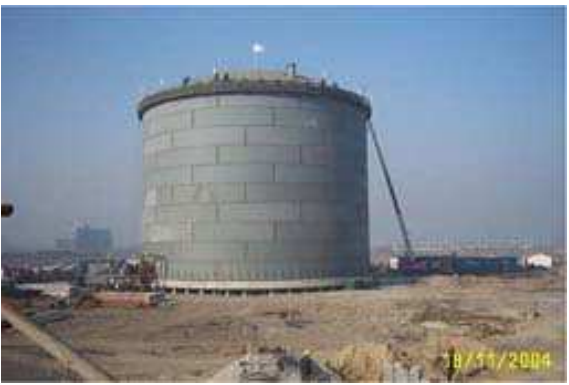,width=7.5cm}
\vspace{-1cm}
\caption{\small A typical 10 kton tanker for LNG storage.}
\label{fig:10ktontanker}
\end{figure}

From the above design considerations, it is clear to us that the realization of a very large LAr TPC will require a graded
program, likely implying the construction of a $\sim$ 1 kton engineering module to assess and implement the adopted technological
and engineering solutions. At a later stage, one should aim at the realization of a $\sim$ 10 kton detector with
a rich physics program on its own. Only after the successful meeting of these milestones, the following factor of 10 mass increase
could be realistically afforded. This step-by-step procedure is justified both from the technological and the financial point of view.
The 10 kton detector would feature a tanker with a diameter of 30 m and a height of 10 m, very similar to the LNG tanker 
shown in Fig.~\ref{fig:10ktontanker}. The number of charge imaging channels will be about 30000. 

A final remark on the detector cost. From our initial estimates the cost of a 100 kton (10 kton) detector built underground
would be around 350 (80) million Euro. 

\section {A 100 ton LAr TPC for the T2K neutrino experiment}

A 100 ton detector in a near site of a long baseline facility is a straightforward and
very desirable application of the LAr TPC technique~\cite{nusag}. 
In particular, the approved T2K experiment in Japan~\cite{Itow:2001ee}
will provide the ideal conditions to study with high statistical
accuracy neutrino interactions on liquid Argon in the very important energy range around 1 GeV.
This is a mandatory step in order to be able to handle high statistics provided by the large detectors
described in the previous Section. In addition, a smaller mass
prototype of the 100 ton detector could offer a tool to study calorimetric (electromagnetic
and hadronic) response in a charged particle beam or be readily placed in an existing neutrino beam.

The J-PARC to Kamioka neutrino project is a second generation long baseline neutrino oscillation 
experiment that will probe physics beyond the Standard Model by high precision measurements of neutrino mixing.
A high intensity, off-axis, narrow band neutrino beam is produced 
by secondary pions created by a high intensity proton synchrotron at J-PARC. 
The neutrino energy is tuned to the oscillation maximum at $\sim$1~GeV for a baseline length of 295~km towards the
Super-Kamiokande detector.

The project is divided into two phases.  In the first, the main goal 
is the precision measurement of neutrino oscillation with the 50~GeV~PS of 
0.75~MW beam power and Super-Kamiokande as a far detector.
The physics subject of the first phase is  an order of magnitude better precision in the $\nu_\mu\rightarrow \nu_x$ 
disappearance oscillation 
measurement, a factor of 20 more sensitive search in the $\nu_\mu\rightarrow\nu_e$
appearance, and a confirmation of the $\nu_\mu\rightarrow \nu_\tau$ oscillation
or discovery of sterile neutrinos by detecting neutral current events. During the second phase, the power of the neutrino
beam will be increased and a new far detector will be built~\cite{Itow:2001ee}.

In order to achieve the challenging goals of the T2K program, for the disappearance oscillation search 
one will need to precisely measure the $\Delta m_{23}^2$ and $\sin^22\theta_{23}$
parameters with small systematic errors. For that, a very good knowledge of the neutrino beam will have to be reached.
For the electron appearance experiment, in order to deeply understand the beam associated backgrounds, 
a very good knowledge of the intrinsic $\nu_e$ component of the beam and of
the $\pi^0$ production in neutrino interactions in the GeV range will be mandatory.
The above requirements are to be fulfilled by the use of a complex detector configuration.
In fact, the T2K long-baseline program foresees two near detector stations, respectively at 140 and 280~m from the target, 
a third intermediate station at 2~km, and the far station represented by the existing Super-Kamiokande detector.
The near detectors have to be built, and will be composed of different technologies, like in the
case of the previous K2K experiment~\cite{Ahn:2002up}.

As far as the 2 km site is concerned, it is now planned to install
a 1~kton water Cerenkov detector, a muon ranger and a so-called "fine grain" detector constituted by
a LAr TPC~\cite{nusag}.
At the 2~km position, the rate in a 100~ton detector would be about 300000 events per
year: this is a unique location for a liquid Argon TPC of such a mass. 
A list of physics measurements that could be performed in the
T2K 2 km near station with a liquid Argon detector has been outlined in~\cite{nuint04}. 
The merits of the detector for the T2K oscillation measurements combined to the other detectors of the 
2 km site are discussed in detail in~\cite{nusag}.  The main technological and detector issues of the 100 ton LAr TPC
for T2K are reported in the following.

\begin{figure}[htb]
\centering
\epsfig{file=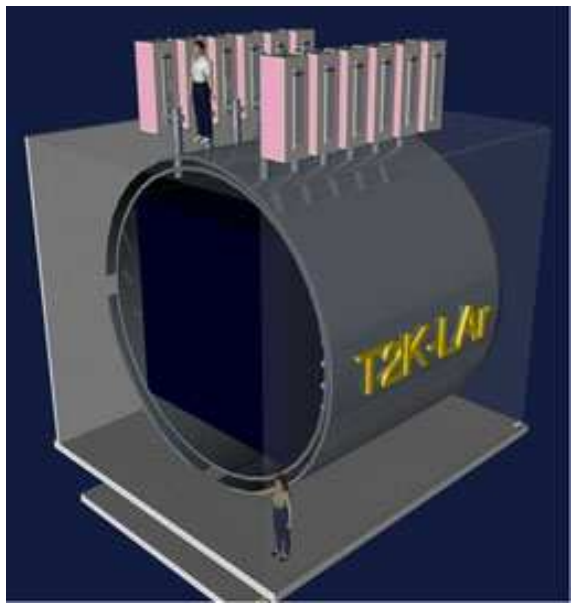,width=7.5cm}
\vspace{-1cm}
\caption{\small Artistic view of the T2K LAr detector vessel.}
\label{fig:T2KLAr1}
\end{figure}

The LAr TPC for the T2K 2 km site is hosted in a 8.5 m long and 7.2 m diameter stainless steel
dewar positioned on mechanical shock absorbers, as schematically shown in Fig.~\ref{fig:T2KLAr1}.
Inside the outer dewar, an inner vessel of 5 m in length and 6.6 m in diameter contains the liquid Argon. 
The volume between the two vessels is evacuated and filled with super-insulation layers to ensure adequate thermal insulation.
The outer cylinder acts as a thin ÒskinÓ for vacuum insulation. The heat input through
these surfaces is estimated to about 100 W under high vacuum conditions. In case of
loss of vacuum, the heat input increases to 4 kW.
The inner vessel contains about 315 tons of liquid. A smaller volume of 150 tons is confined by the
inner TPC, corresponding to a neutrino interaction fiducial volume of 100 tons. 
The basic concept for the liquid Argon cryostat has been developed and engineered 
following the internationally recognized codes for the design of conventional cryogenic-fluid
pressure storage-vessels as covered in the ASME (American Standards of Mechanical Engineers) Boiler \&
Pressure Vessel Code, Sect. VIII. 

The detector chamber consists of a stainless steel mechanical frame with parallelepiped shape inscribed
in the inner vessel cylinder.
The cathode of the TPC is placed in the middle of the inner volume along the longitudinal axis. There
are two options for this element: either filled with frozen water or with solid CO$_2$.  The inner target is motivated by the fact
that the extrapolation between Argon and water targets (2 km water Cerenkov and Super-Kamiokande)
can be affected by uncertainties, which in turn would affect
the goal of precision measurement of the oscillation parameters. 
The straightforward solution is, therefore, to insert an additional target within the 100 ton liquid Argon detector
in ordere to collect a statistically significant event sample with interactions occurring in Oxygen (CO$_2$ or H$_2$O)
and tracks reconstructed in LAr.

\begin{figure}[htb]
\centering
\epsfig{file=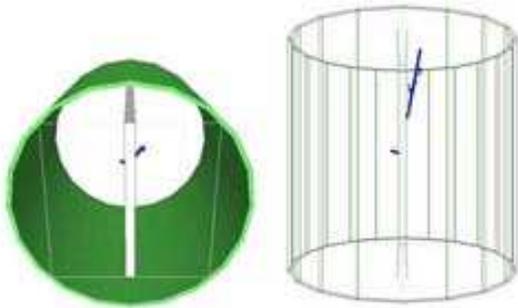,width=7.5cm}
\vspace{-1cm}
\caption{\small The inner dewar volume with the option of a
parallelepiped shaped inner target. A simulated neutrino event is also shown.}
\label{fig:MCwater}
\end{figure}

\begin{figure}[htb]
\centering
\epsfig{file=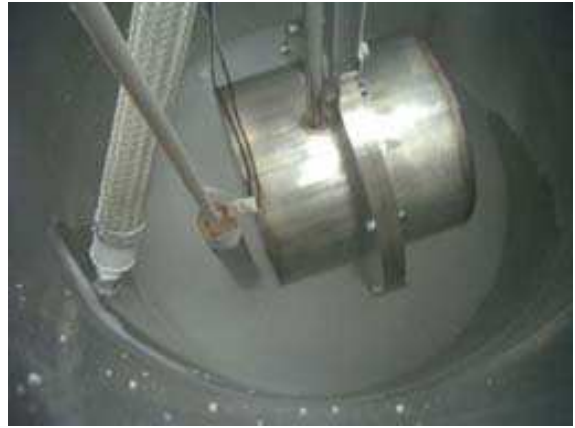,width=7.5cm}
\vspace{-1cm}
\caption{\small A prototype cylindrical water target during a freezing test with liquid Argon.}
\label{fig:targetprototype}
\end{figure}

The first option for the inner target geometry
is a cylindrical structure made of a 2 mm thick stainless steel cylinder of 60 cm diameter
and 5 m length. The second option calls for a parallelepiped shaped target
25 cm thick and 5 m long  (Fig.~\ref{fig:MCwater}). In the latter case, two separate cathode planes would be placed onto
the external sides of the target. For both options, the cathode electrode defines two half-volumes. 
We are conducting specific tests with prototype targets to study the freezing procedure of the target material.
Fig.~\ref{fig:targetprototype} shows one of these (cylindrical) prototypes during a laboratory test to 
simulate water target freezing inside a dewar being filled with liquid Argon.

Each of the two extreme sides of the half-volumes are equipped with two or three wire planes with different wire
orientations, constituting the readout anodes.
The electric field perpendicular to the wires is established in the LAr volume by means of a high voltage (HV)
system. The system is composed
of the above-mentioned cathode plane parallel to the wire planes, placed in the center of the cryostat
volume at a distance of about 2 m from the wires of each side and defining the maximum drift length, and of
field shaping electrodes made of stainless steel tubes. These are required to guarantee
the uniformity of the field along the drift direction. At the nominal cathodic voltage of about 200 kV,
corresponding to an electric field of 1 kV/cm, the maximum drift time in LAr is of about 1 ms. 
At this stage, two options are considered for polarization of the cathode and the corresponding electrodes. In
the first option a HV feedthrough allows to set the required potential on the cathode. The electrodes are placed
at degrading voltage from the cathode potential to ground via a series of resistors inserted between the
tubes.  In the second option, the use of a HV feedthrough is avoided if an oscillating low-voltage
is multiplied inside the detector via a Greinacher circuit (see next Section).

Large area PMTs are placed inside the liquid, attached to the supporting mechanical structure,
outside the inner fiducial volume and behind the wire planes. The PMTs are manufactured to be sensitive
to the DUV prompt scintillation of Argon. These signals can be used for triggering on non beam associated events. 

On top of the cryostat there are flanges equipped with cryogenic feedthroughs for the electrical connection
of the wires with the readout electronics. These feedthroughs also provide passage for the internal
instrumentation including PMTs, purity monitors, level meters, temperature probes, etc. The electronics
allows for continuous readout, digitization and wave-form recording of the signals from each wire of the
TPC. The frontend electronics is hosted in crates directly placed on top of the dewar.

The detector is complemented by ancillary cryogenics systems. A gas and liquid recirculation and
purification system, a heat exchanger and a LAr buffer are placed in the underground cavern close to the
detector dewar. These systems are connected through cryogenic pipes to the surface, where Argon storage,
compressor, ventilation and evaporation systems complete the detector infrastructure.

\begin{figure}[htb]
\centering
\epsfig{file=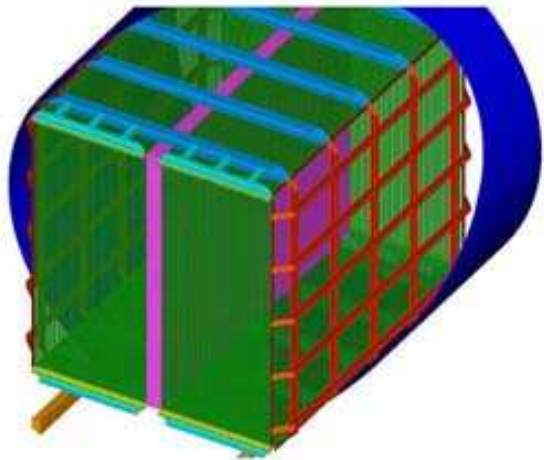,width=7.5cm}
\vspace{-1cm}
\caption{\small Layout of the inner detector supporting structure. One can notice the central inner target, the race-tracks,
the sustaining rails and one of the two chamber supporting frames.}
\label{fig:TPC}
\end{figure}

In more detail, the inner detector is composed of a rigid mechanical frame made of stainless-steel beams that is based on
the cryostat floor by means of a system of rails and adjustable feet. The structure is made independent
from possible deformations of the cryostat occurring during the initial cooling and it is also self-supporting.
The stainless-steel frame has dimensions of about 5 m in length, 4.5 m in width and 4.5 m in height  (Fig.~\ref{fig:TPC}).
Two lateral wire-frames with the task of supporting the two (left and right) TPCs are positioned on the
vertical longitudinal sides of the mechanical structure. The structure is designed in order to sustain the
total force applied by the TPC wire tension without appreciable deformations. In the baseline option of two wire planes
per each detector half this force amounts to about 9 ton.
The structure is also designed to hold the signal and HV connectors, the PMTs, and the monitoring
devices. Once assembled, the structure equipped with TPC planes and the other devices will be moved
into the dewar by means of the bottom rails and subsequently aligned and placed in its nominal position.
 
As mentioned above, each TPC consists of a system of two parallel wire planes separated by 3 mm. 
The wire pitch normal to the wire direction is 3 mm. An additional
third wire plane in each chamber would increase redundancy in the event reconstruction and help
in solving possible ambiguities.  Wires are made of AISI 304V stainless steel with a diameter of 150 $\mu$m. 
They are stretched in the frame
supported by the mechanical structure previously described. Each of the two-wire-plane chambers contain
about 4500 wires, some of which decreasing in length. The tension applied to each wire
corresponds to 10 N in the baseline option, although some optimization is expected following laboratory test we are 
presently conducting.
 
It is worth to stress a few points concerning the detector operation.
Since no active refrigeration is foreseen, there are in total about 40 tons of steel to be cooled
by liquid Argon boil-off during the initial cooling. 
The amount of Argon needed for this is about 20 LAr m$^3$, less than 10\%
of the total stored Argon (315 ton). 
In normal operating conditions the total heat input induces a boil-off of about 200 l of liquid Argon
per day, that is quite small. During forced liquid
recirculation, required to yield high purity,  the consumption is estimated to increase by additional 400 l/day.
Since the cryostat is not actively cooled, $e.g.$ with liquid N2, the cold input is externally provided either
by compensation of the boil-off or preferably via a standard Linde-Hampson refrigeration process.

Refrigeration of the liquid Argon volume is performed by means of a closed circuit with a compressor
placed at the surface, operated by a feed-back on the temperature and pressure of the inner vessel. Heat
losses under normal operating conditions correspond to 500 W (cold) or 10 kW of electric power. If we
include the power required for the liquid Argon recirculation/purification we
reach a total of about 30 kW of electric power to be supplied.

The required LAr purity is firstly ensured by using suitable materials, cleaning, and careful design
of the internal components. Additionally, internal surfaces can be vacuum conditioned, since during
steady operation pollution of the LAr is mainly due to out-gassing of the inner surfaces and inner detector
materials in contact with the gaseous Argon.
The second requirement is that filling of the detector dewar from the external storage must be performed
through sets of filters placed in series. Each set is dimensioned to allow for
the purification of the LAr volume starting from standard commercial LAr (with a concentration of water and Oxygen 
of about 0.5 ppm).
The third requirement is that a recirculation system has to be implemented in order to reach the ultimate
purity of the liquid at equilibrium during detector operation, compensating for sources of impurities,
typically degassing of materials immersed in the liquid. During the refrigeration cycle
the gaseous Argon is passed through a purification filter before it is filled
back as LAr to the inner vessel. Additional Argon purification is provided by a recirculation system
through purification cartridges in liquid phase.  A similar system turned out
to be very important in the ICARUS T600 operation~\cite{gg3}.
The liquid recirculation system can be dimensioned under the requirement of a 48 hours recirculation
period. This implies a recirculation rate of about 3500 LAr l/hour (1 l/s). This can be
handled by employing 7-8 filter cartridges arranged in two parallel circuits.

More details about the LAr TPC for T2K can be found in~\cite{nusag}.

\section{R\&D towards a large mass liquid Argon TPC}

Some studies are underway and others are planned with the aim of optimizing the design of future large mass LAr TPC
detectors~\cite{multimw,nufact04}. We report here on the general R\&D strategy and on 
results we have recently achieved.

The development of suitable charge extraction, amplification and collection devices is a crucial issue.
We are continuing an R\&D activity to further optimize the technique for charge extraction, amplification and collection,
seeking a solution which yields gains between 100 and 1000 in pure Argon, that is electrically
and mechanically stable, and easy to be mass produced.
The starting point can be the Gas Electron Multiplier (GEM) detector developed by F.~Sauli and coworkers~\cite{Sauli:qp}.
It consists of a thin, metal-clad polymer foil, chemically pierced by a high density of holes. 
On application of a difference of potential between the two electrodes, 
electrons released by radiation in the gas on one side of the structure drift into the holes, 
multiply and transfer to a collection region.Ê

A detector derived from the GEM on which we are presently conducting specific studies is
the so called LEM (Large Electron Multiplier)~\cite{Jeanneret:mr}. It can be considered as a sort of macroscopic GEM
built with a thick vetronite-Cu coated board (1-2 mm) with relatively  large holes of 0.5 mm diameter.
These features might make more easy the operation at cryogenic temperatures.

\begin{figure}[htb]
\centering
\epsfig{file=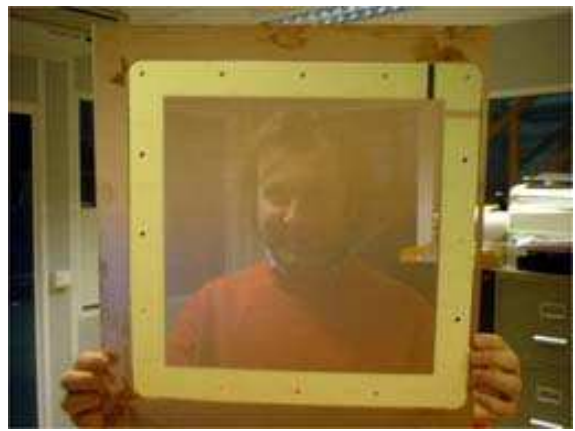,width=7.5cm}
\vspace{-1cm}
\caption{\small A prototype LEM detector used in the test measurements.}
\label{fig:LEM}
\end{figure}

The LEM detector is shown in Fig.~\ref{fig:LEM}. In our preliminary measurements we have been studying
the gain yield as a function of HV and gas pressure.  
Gains up to 800 seem to be achievable even at high pressure, with good prospects for operation in cold.
A preliminary resolution of about 28\% FWHM has been obtained for a Fe source. The 
experimental results agree with those expected from simulations.

\begin{figure}[htb]
\centering
\epsfig{file=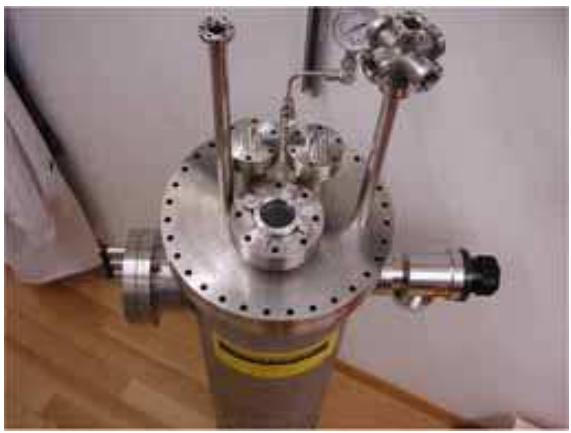,width=7.5cm}
\vspace{-1cm}
\caption{\small The vessel housing the detector to study operation at high hydrostatic pressure.}
\label{fig:highpressure}
\end{figure}

We are also conducting studies for
the understanding of charge collection under high pressure as occurring for events occurring at the bottom 
of the large cryogenic tanker.
At this purpose, we constructed a small chamber (Fig.~\ref{fig:highpressure}) which will be pressurized to 3-4~bar to simulate
the hydrostatic pressure at the bottom of a future 100~kton tanker. We plan
to check that the drift properties of electrons are actually not affected at these pressure values.

Another important subject of our current R\&D studies is the problem of delivering very HV to the
inner detectors, trying to avoid the use of (delicate) high voltage feedthroughs.
We have realized a series of device prototypes based on the Greinacher or Cockroft-Walton circuit that allows to 
bring into the vessel a relatively low voltage and operate the required amplification directly inside the cryogenic liquid.

\begin{figure}[htb]
\centering
\epsfig{file=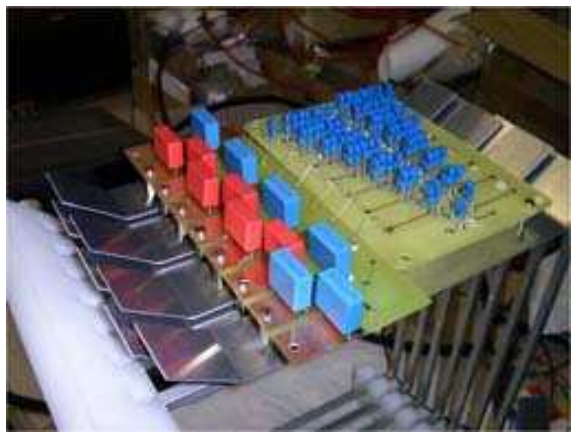,width=7.5cm}
\vspace{-1cm}
\caption{\small One of the prototype Greinacher circuits used for the tests.}
\label{fig:greinacher}
\end{figure}

\begin{figure}[htb]
\centering
\epsfig{file=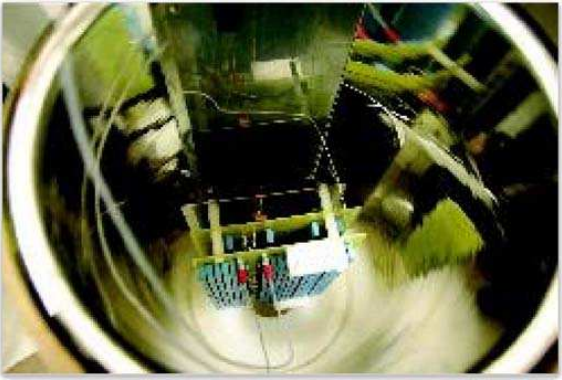,width=7.5cm}
\vspace{-1cm}
\caption{\small The Greinacher circuits during the "cold" tests with liquid Nitrogen.}
\label{fig:greinacher2}
\end{figure}

In preliminary tests with 20 amplification cells we have been able to reach about 40 kV. Nearly 2 kV/cm were obtained
with the circuit successfully operating in liquid Nitrogen. A filter was built with two RC circuits in series
in order to attenuate the ripple of the DC output by about 20 dB. 
The measurements indicate a very low
noise level induced to the readout acquisition channels. However, in order to solve the problem of noise
generation by the AC input signal of the HV circuit, some other solution can be envisaged. We foresee
the use of an input signal at 50 Hz, a frequency very far from the bandwidth of any commercial preamplifier
that can be used for the wire readout. A second possibility is to stop the AC signal when an event trigger
is issued, as long as the full drift and acquisition time last.
Further tests to reach 200 kV are ongoing and long term stability studies in cold are foreseen.
Fig.~\ref{fig:greinacher} shows the implementation of one of the Greinacher circuits used for the test measurements.
Fig.~\ref{fig:greinacher2} illustrates the operation of the circuit immersed in liquid Nitrogen.

\begin{figure}[htb]
\centering
\epsfig{file=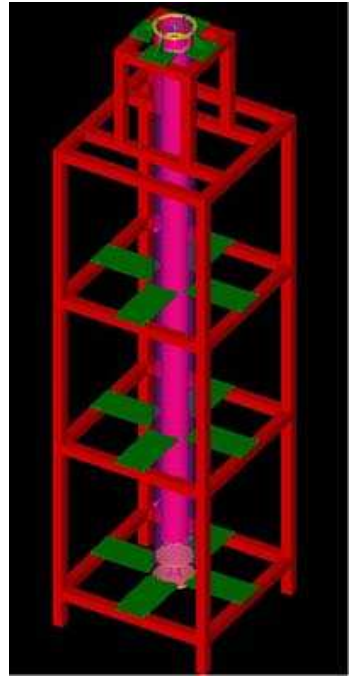,width=4.5cm}
\vspace{-1cm}
\caption{\small Engineering design of the ARGONTUBE prototype LAr TPC.}
\label{fig:argontubedesign}
\end{figure}

\begin{figure}[htb]
\centering
\epsfig{file=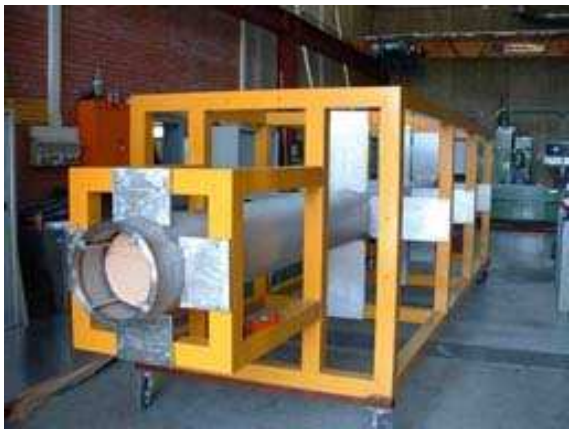,width=7.5cm}
\vspace{-1cm}
\caption{\small The long detector column after assembly in the support structure.}
\label{fig:inthesupport1}
\end{figure}

\begin{figure}[htb]
\centering
\epsfig{file=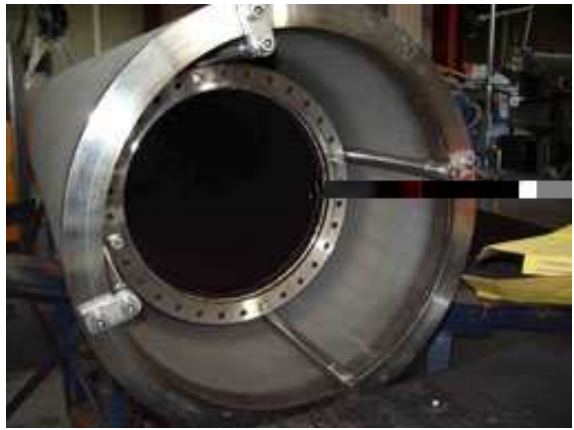,width=7.5cm}
\vspace{-1cm}
\caption{\small Detail of the inner ARGONTUBE detector inside the dewar vessel.}
\label{fig:detailargontube}
\end{figure}

The realization of a 5 m long detector column will allow to experimentally prove the feasibility of detectors with long drift path
and will represent a very important milestone in our global strategy.
We have recently designed (Fig.~\ref{fig:argontubedesign})
and completed the construction of a 6 m long and 40 cm in diameter dewar 
vessel housing a 5 m long prototype LAr detector (ARGONTUBE). The device will be operated 
with a reduced electric field value in order to simulate very long drift distances of up to 20 m.
Charge attenuation and amplification will be studied in detail together with the adoption of possible novel technological
solutions. We also plan to implement a high voltage system based on the previously described Greinacher approach.
Fig.~\ref{fig:inthesupport1} shows the 5 m long TPC tube inserted in its supporting structure, while
Fig.~\ref{fig:detailargontube} depicts a detail of the inner detector tube inside the dewar vessel.

The study of LAr TPC prototypes in a magnetic field is an important breakthrough of the technique.
Liquid Argon imaging provides very good tracking with $dE/dx$ measurement, and excellent calorimetric 
performance for contained showers. This allows for a very
precise determination of the energy of the particles in
an event, in particular for electron showers, which energy is very precisely measured. 
The possibility to complement these features with those provided by 
a magnetic field has been considered~\cite{Rubbia:2001pk,Rubbia:2004tz}  
and would open new possibilities such as
charge discrimination, momentum measurement of particles escaping the detector, and
very precise kinematics, since the measurements are multiple scattering
dominated ($e.g.$ $\Delta p/p\simeq 4\%$ for a track length of 12 m and a B field of 1T).

The orientation of the magnetic field is such that the bending is in the direction of the drift, where the 
best spatial resolution is achieved (in the ICARUS T600 a point resolution of $400\ \mu m$ was obtained).
The magnetic field is hence perpendicular to the electric field. The Lorentz angle is expected to be very small in 
liquid. Embedding the volume of Argon into a magnetic field would therefore not
alter the imaging properties of the detector and the measurement of the bending of
charged hadrons or penetrating muons would allow a precise determination of the momentum and a determination of their charge.
For muons, a field of 0.1 T allows to discriminate the charge for tracks longer than 4 m,
corresponding to a momentum threshold of 800~MeV/c.
Unlike muons or hadrons, the early showering of electrons makes their charge identification difficult. 
From simulations it is found that the determination of the charge of electrons of energy in the range between
1 and 5 GeV is feasible with good purity, provided that the field has a strength in the range of 1~T.

\begin{figure}[htb]
\centering
\epsfig{file=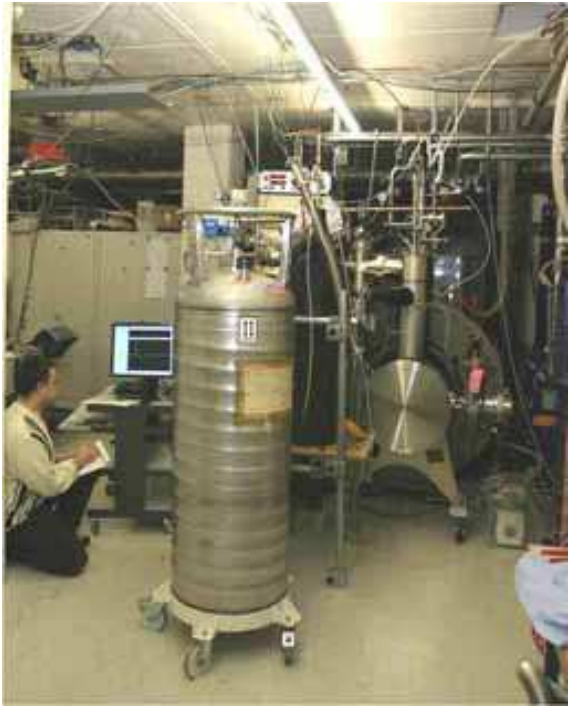,width=7.5cm}
\vspace{-1cm}
\caption{\small The setup used for the measurements of LAr TPC operation 
in magnetic field at ETHZ Zurich.}
\label{fig:magneticfield}
\end{figure}

\begin{figure}[htb]
\centering
\epsfig{file=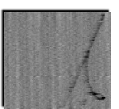,width=3.8cm}
\epsfig{file=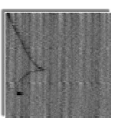,width=3.5cm}
\vspace{-1cm}
\caption{\small Cosmic-ray events taken with the LAr TPC detector in magnetic field.}
\label{fig:event}
\end{figure}

An R\&D program to study the performance of a LAr TPC in a magnetic field was started in 2001. 
For this purpose, at ETHZ Zurich we have built a small liquid Argon TPC (width 300~mm, height 150~mm, drift
length 150~mm) and placed it in the SINDRUM-I magnet kindly provided by the PSI Laboratory, producing
field intensities up to 0.5~T (Fig.~\ref{fig:magneticfield}). The test program was conducted successfully 
and included checking the basic imaging, measuring traversing and stopping muons, testing charge
discrimination, and checking of the Lorentz angle. The results have been recently published~\cite{laffranchi}.
Two cosmic events collected with the device are shown in Fig.~\ref{fig:event}.

The further development of the industrial design of a large volume tanker able to operate underground will be
also pursued in the framework of our future activities.
The study initiated with Technodyne UK should be considered as a first ``feasibility'' study meant to 
select the main issues that will need to be further understood and to promptly identify possible ``show-stoppers''.
We expect to continue this study by more elaborated and detailed industrial design of the
large underground (or shallow depth) tanker also including the details of the detector instrumentation. 
As anticipated, 
the cost of the full device will be estimated in more detail as well. At this preliminary stage a large mass LAr detector
does appear to be a cost effective option.

Finally, we are investigating the study of logistics, infrastructure and safety issues related to underground sites.
We are making investigations with two typical geographical configurations: a tunnel-access underground
laboratory and a vertical mine-type-access underground laboratory. Early considerations show that such sites correspond
to interesting complementary options.
Concerning the provision of LAr, a dedicated, possibly not underground but nearby, air-liquefaction 
plant is foreseen. In collaboration with Technodyne 
we have started addressing the technical requirements and feasibility of such a facility.

\section{Conclusions and outlook}

The basic R\&D work for the liquid Argon Time Projection Chamber technology has been successfully conducted 
by the ICARUS Collaboration, proving that it represents
a mature particle detector technique with great potential for future neutrino and astroparticle
physics experiments~\cite{t600paper}.  

In the last few years we have outlined a global strategy for next generation 
experiments based on the use of this technique at different detector mass scales~\cite{nuint04}. 
Our conceptual design of a 100 kton liquid Argon TPC seems technically sound;
if realized, this detector would deliver extraordinary physics output and 
effectively complement giant 0.5-1 Megaton water Cerenkov detectors being proposed for 
future precision studies of the neutrino mixing matrix and for nucleon decay searches~\cite{nakamura,kajita}.
Coupled to future Super Beams, Beta Beams or Neutrino Factories
it could greatly improve our understanding of the mixing matrix in the lepton sector with
the goal of measuring the leptonic CP-phase, and in parallel it would
allow to conduct astroparticle experiments of unprecedented sensitivity.

The main design features of a large LAr TPC include the possibility of
a double-phase operation with charge amplification for long drift distances,
a charge imaging plus light readout for improved physics performance, and
a very large boiling industrial cryostat (LNG technology).
A full-scale, cost effective "prototype" at the scale of
10 kton could be envisaged as an engineering design test with a physics program
on its own.  A first module with a mass of ~1 kton could 
effectively play the role of a "module zero" meant to assess the main technological issues
of the design, its scalability to larger masses, and to verify the solutions adopted after 
the R\&D studies conducted on specific subjects.

A 100 ton LAr TPC detector in a near-site of a long-baseline facility is a straightforward and
desirable application of the technique~\cite{nuint04}. The T2K experiment presently under construction in Japan
will provide the ideal conditions to study with high statistical
accuracy neutrino interactions on liquid Argon in the very important energy range around 1 GeV~\cite{Itow:2001ee}.
A proposed detector placed in the T2K 2 km site could provide the required reduction of the systematic errors
for a high sensitivity measurement of the so far unknown $\theta_{13}$ mixing angle~\cite{nusag}. 
Also in this case,
a smaller mass prototype would represent a useful tool to study calorimetric (electromagnetic
and hadronic) response in a charged particle beam.

Work is in progress along the above lines of thoughts, with undergoing design and optimization studies, as well as
with specific technical R\&D activities.  There is a high degree of interplay and a strong
synergy between small and large mass scale apparatuses, the very large
detector needing the small one in order to best exploit the measurements with high statistical precision
that will be possible with a large mass. 

In conclusions, we do believe that the liquid Argon TPC technique is one of the favorite detector approaches
to meet the challenge of future experiments both on neutrino and on astroparticle physics. From past experience and undergoing
studies one can be confident that the detector mass could actually scale from the 100 ton to the 100 kton mass range 
while keeping the basic outstanding features unaffected, and hence allowing for the design and the construction of 
cost effective and reliable devices, suitable for long term operation.

\section{Acknowledgments}
We kindly acknowledge the work of all the colleagues contributing to the data, to the experiments, and to
the design and R\&D activities presented in this paper.
A.E. wishes to warmly thank F. Cervelli for the kind invitation to HIF05 and for the excellent organization of the workshop.

\end{document}